\documentclass[conference]{IEEEtran}
\IEEEoverridecommandlockouts
\usepackage{cite}
\usepackage{amsmath,amssymb,amsfonts}
\usepackage{algorithmic}
\usepackage{graphicx}
\usepackage{textcomp}
\usepackage{xcolor}
\usepackage{tikz}
\usepackage{verbatim}
\usepackage{adjustbox}
\usepackage{multirow}
\usepackage{enumitem}
\usepackage{hyperref}
\usepackage{booktabs}
\def\BibTeX{{\rm B\kern-.05em{\sc i\kern-.025em b}\kern-.08em
    T\kern-.1667em\lower.7ex\hbox{E}\kern-.125emX}}

\usepackage[english]{babel}
\addto\extrasenglish{%

}

\begin{document}

\title{Towards AI-enabled Cyber Threat Assessment in the Health Sector\\}

\author{\IEEEauthorblockN{1\textsuperscript{st} Patrizia Heinl}
\IEEEauthorblockA{\textit{Faculty of Computer Science} \\
\textit{Technische Hochschule Ingolstadt}\\
Ingolstadt, Germany \\
patrizia.heinl@thi.de}
\and
\IEEEauthorblockN{2\textsuperscript{nd} Andrius Patapovas}
\IEEEauthorblockA{\textit{Faculty of Computer Science} \\
\textit{Technische Hochschule Nürnberg}\\
Nürnberg, Germany \\
andrius.patapovas@th-nürnberg.de}
\and
\IEEEauthorblockN{3\textsuperscript{rd} Michael Pilgermann}
\IEEEauthorblockA{\textit{Faculty of Computer Science} \\
\textit{Technische Hochschule Brandenburg}\\
Brandenburg an der Havel, Germany \\
michael.pilgermann@th-brandenburg.de}
}
\maketitle

\begin{abstract}
Cyber attacks on the healthcare industry can have tremendous consequences and the attack surface expands continuously. In order to handle the steadily rising workload, an expanding amount of analog processes in healthcare institutions is digitized. Despite regulations becoming stricter, not all existing infrastructure is sufficiently protected against cyber attacks. With an increasing number of devices and digital processes, the system and network landscape becomes more complex and harder to manage and therefore also more difficult to protect. The aim of this project is to introduce an AI-enabled platform that collects security relevant information from the outside of a health organization, analyzes it, delivers a risk score and supports decision makers in healthcare institutions to optimize investment choices for security measures. Therefore, an architecture of such a platform is designed, relevant information sources are identified, and AI methods for relevant data collection, selection, and risk scoring are explored.
\end{abstract}


\section{Introduction}

At a time when digitization is permeating all aspects of our lives, ensuring digital security in the healthcare sector is becoming an urgent imperative. This becomes even more crucial as, on the one hand, life expectancy rises, leading to a surge in diseases necessitating treatment. On the other hand, patient data that is particularly worthy of protection and the essential role of healthcare in our society demand a robust security framework.
Up to date numbers speak for themselves: Researchers found that in the time from 2016 to 2022, 6,835 healthcare companies in the US were hit by ransomware~\cite{weber_attack_2023}. A reply of the German federal government to the Federal Parliament revealed, that in the time from 2018 to 2023, 224 IT security incidents were reported from those hospitals, which are identified as critical infrastructures based on the IT security legislation~\cite{die_bundesregierung_antwort_2024}. 

For making critical infrastructures such as hospitals cyber secure, the security teams heavily depend on information about the threats, the organizations face. However, in order to generate a real added value, multiple requirements have to be met. Relevant cyber threat intelligence must be accessible, correct, trustworthy and up to date. Therefore, technical scans of organizations' infrastructure have to be performed and information sources in the public domain have to be identified. Once identified, data also has to be processed, assessed and integrated for a comprehensive perspective on the situation. As IT always serves the business, the potential impact of threats on business processes of the organization has to be considered to increase the meaningfulness of assessments drastically. This publication faces those challenges by introducing an AI-enabled platform to continuously monitor and assess the cyber risk of the health organizations. The novel architecture of the platform is introduced in \autoref{label_platformarchitectur}. The following chapters, \autoref{label_deepdive1} and \autoref{label_deepdive2}, focus on the input data of the platform while \autoref{label_ProcessMapping} sheds light on how the input data is mapped to the healthcare business processes. Finally, the last chapter, \autoref{label_security_scoring} introduces the concept of an AI-enabled risk scoring model.

\section{Platform AI4HCTI}
\label{label_platformarchitectur}

The envisioned AI-enabled platform for IT security information exchange within a healthcare network aims to achieve three primary objectives:
\begin{itemize}
    \item to enhance the data model for comprehensive monitoring and cyber risk assessment of the healthcare organization using advanced algorithms,
    \item to integrate Cyber Threat Intelligence (CTI) to improve communication between IT security and operational teams as well as
    \item to efficiently systematize healthcare organizations’ insights for early detection of cyber attacks and to initiate countermeasures.
\end{itemize}

Additionally, it provides users with AI-enabled recommendations based on analyzed data about vulnerabilities, threats, and attacks, aiming to proactively protect critical infrastructure. In comparison to earlier approaches to build a platform for a NLP-based cyber threat assessment \cite{lakka2022incident}\cite{silvestri2023machine} or active incident handling \cite{papastergiou2019cyber}\cite{papastergiou2021handling}, the AI-enabled Health Cyber Threat Intelligence (AI4HCTI)-platform does not implicate a conduction of local agents at healthcare organizations to automatically aggregate incident handling at the platform. Cyber strategies such as \textit{zero trust}, as the antithesis of the traditional adage of \textit{“trust but verify"}, suggest to separate the control plane from external sources and uses trust algorithms to decide resource access to the network\cite{scalco2021control}. Furthermore, a systematic review of 2577 articles by Walker-Roberts et al. (2017) recommends to make all resources in a healthcare critical infrastructure immutable to attackers while limiting data access and permission available to insiders \cite{walker2018systematic}.

\subsection{General Architecture}
\begin{figure}[h]
  \centering
  \includegraphics[width=\linewidth]{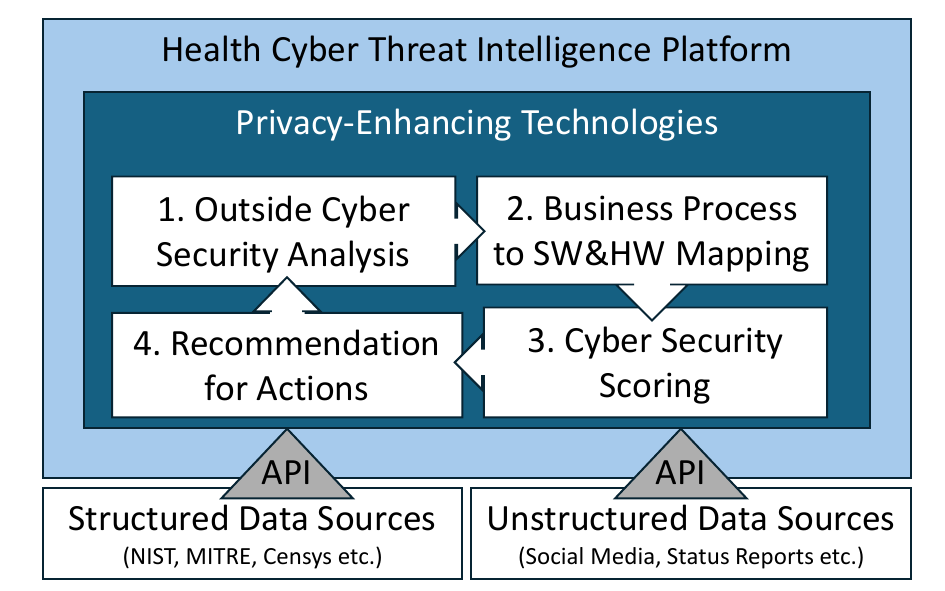}
  \caption{Core components of AI-enabled HCTI platform.}
  \label{fig:AI4HCTI_Components}
\end{figure}

The AI4HCTI-platform, designed to enhance cybersecurity in healthcare organizations, consists of several integrated components and layers (see \autoref{fig:AI4HCTI_Components}). The platform does not only assess external cybersecurity information of an organization but also closely maps these evaluations to the specific business, software, and hardware layers within the organization. The ultimate goal is to provide actionable insights and recommendations to fortify security measures. Below is a breakdown of the general platform architecture:

\begin{enumerate}[wide, labelwidth=!, labelindent=0pt]
    \item \textit{Data collection and outside cybersecurity scoring}: This component gathers data from external sources to evaluate the cybersecurity posture of the organization. It may include reports from cybersecurity rating agencies, threat intelligence feeds, and other relevant external data sources. This scoring helps to benchmark the organization against industry standards and peer institutions. In addition, it closes the existing lack of datasets in cyber risks assessment \cite{cremer2022cyber}. 
    \item \textit{Enterprise architecture mapping}: Existing enterprise architecture models help navigating complex business processes and can be used to enhance security strategies \cite{jiang2024enterprise}. This involves mapping out a relevant collection of critical business processes, particularly those that handle sensitive medical data or that are relevant for security of supply. The mapping allows to identifying which software applications and hardware devices support the critical processes. The view on enterprise architecture ensures that any cybersecurity measures taken can be accurately aligned with the specific needs of different business units and their respective technologies. Additionally, the implementation of rapid technological advancement and evolving federal policy, which are the main drivers  exposing healthcare to cyber threats \cite{kruse2017cybersecurity}, is more comprehensible. 
    \item \textit{Cybersecurity Scoring}: Using AI algorithms, the platform analyzes the collected data to detect anomalies, signs of potential cyber threats, or breaches in the context of the current threat landscape and the specific vulnerabilities of the healthcare organization. This analysis includes pattern recognition, predictive analytics, and behavioral analysis to proactively identify risks. Leveraging AI, the system adapts its scoring and detection algorithms based on new data and emerging threats, ensuring that the organization’s defenses evolve as needed.
    \item \textit{Recommendations and actionable insights}: Based on the outside and inside scoring, the platform generates specific recommendations for improving cybersecurity measures. These recommendations are tailored to address the identified vulnerabilities at the organizational, application, and technological layers. A dashboard provides stakeholders with an overview of the cybersecurity health of the organization, detailing threats, vulnerabilities, and actions proposed as well as monitoring based on the enterprise archtecture mapping.
\end{enumerate}

\subsection{Privacy-enhancing Technologies}
Privacy-enhancing technologies (PET) are tools that support the protection of sensitive information while allowing to further use the data and run calculations on it~\cite{PET_RoyalSociety}. As the AI4HCTI-platform collects and analyzes sensitive security information of healthcare institutions, privacy enhancing technologies need to be considered to prevent malicious actors from misusing the information. Examples of PET-concepts, such as differential privacy and federated learning, as defined in~\cite{PET_RoyalSociety} are assessed with regards to practicability.

\section{Threat intelligence including information about vulnerabilities}
\label{label_deepdive1}

Following Menges \cite{menges_cyber_2020}, a good starting point for defining Cyber Threat Intelligence can be found in a 2016 report of the Bank of England. Therefore, Cyber Threat Intelligence can be considered as "Information about threats and threat actors that provides sufficient understanding for mitigating a harmful event" \cite{bank_of_england_cbest_2016}. Furthermore, Menges revealed, that existing data structures often only support low-level information, while information on a semantically higher level cannot be represented \cite{menges_cyber_2020}. 

\subsection{Challenges in Health CTI}

The stock take of the threat intelligence situation concretely in healthcare brings out loads of potential for improvement. By law, manufacturers of medical devices must provide information about vulnerabilities of their product to the operators. Besides this, availability of tailored threat intelligence for operators of hospitals has significant limitations:

\begin{itemize}
    \item Open source threat intelligence is available, however it is not tailored to IT environments, which are existent in hospitals. Other parts of hospital infrastructures such as networked medical devices or facility technology is rarely covered at all.
    \item Service providers offer services for supplying threat intelligence; again these services rarely focus on the numerous specialties in hospitals.
    \item Health industry specific threat intelligence is very rare, only in 2023 an initiative from stakeholders in the US named \emph{Health Sector Cybersecurity Coordination Center (HC3)}\footnote{\url{https://www.hhs.gov/hc3}} started providing information, which is directly focusing the challenges in the health industry. Their products include monthly threat briefs and sector alerts.
    \item Although formats and technologies for exchanging threat intelligence have been available for a while, in practice relevant information chunks are often shared between trusted individuals and via e-mail.
\end{itemize}

The authors' approach aims to tackle these limitations in two dimensions: firstly, the relevant CTI technologies and protocols have to be adopted and relevant sources have to be selected to unlock their potential for healthcare environments - this will be described in more detail in \autoref{label_prestructured}. Secondly, an innovative approach has to acknowledge, that relevant sector tailored threat intelligence is currently primarily made available as natural language information. \autoref{label_nlp} will give a preliminary insight on how natural language processing technologies shall be applied in order to provide relevant and up to date threat intelligence to operators of hospitals.

\subsection{Relevant Sources}
\label{label_sources}

In summer 2023 a survey was carried out asking experts from hospitals in Germany about their preparedness regarding detection and handling of IT security incidents. As part of this survey, the participants also provided feedback on which sources they use practically in order to gather relevant threat information.

Although the complete results of this survey are still under finalization and approval as part of the project report, some general conclusions are possible. Regarding the feedback from the experts, the following sources are primarely used as threat intelligence sources:

\begin{itemize}
    \item Web sites of public bodies and their products - in Germany mainly the Federal Office for Information Security, BSI, but also the European Union Agency for Cybersecurity (ENISA),
    \item MISP instances of public bodies as well as open ones such as CIRCLE \footnote{\url{https://www.circl.lu/services/misp-malware-information-sharing-platform/}},
    \item Other sources with pre-structured information such as \emph{abuse.ch},
    \item Service providers specialized on CTI,
    \item Networks of Computer Security Incident Response Teams (CSIRT) or Computer Emergency Response Teams (CERT),
    \item News, blogs and X feeds, as well as
    \item briefs of healthcare specific associations such as the \emph{Hospital Federations} in Germany.
\end{itemize}

Information about vulnerabilities is mainly received from the manufacturers. Also public bodies, media and vulnerability tailored services such as the Industrial Control Systems Cyber Emergency Response Team (ICS-CERT)\footnote{\url{https://www.cisa.gov/topics/industrial-control-systems}} or  CVEDetails\footnote{\url{https://www.cvedetails.com/}} provide additional information in this sense \cite{stine_cyber_2017}. 

\subsection{Integration of Pre-structured Information}
\label{label_prestructured}

Generally speaking, as introduced in \autoref{label_deepdive1}, threat information from the different sources (\autoref{label_sources}) may either be presented in a pre-structured format, or presented in non-structurecd format - mostly natural language documents such as threat briefs or e-mails. For supporting the integration of pre-structured information in a platform like AI4HCTI, the relevant formats must firstly be identified.

\subsubsection{Potentially Relevant Formats}

Menges researched into the exchange of CTI during his PhD and published his research results in a set of papers. The second paper named "\emph{Unifying Cyber Threat Intelligence}" \cite{gritzalis_unifying_2019} analyzed serveral data formats aiming towards standardization of these data structures. 
Taking the results of their research into account, the following exchange formats and data structures for CTI existed in 2020:

\begin{itemize}
    \item Structured Threat Information eXpression 1 / 2 (STIX) \cite{sean_barnum_standardizing_2014} from Mitre, being the most widely used data format in the field of CTI;
    \item Incident Object Description Exchange Format (IODEF) and IODEF2\footnote{ \url{https://tools.ietf.org/html/rfc5070},  \url{https://tools.ietf.org/html/rfc7970}}, focussing on security incident reports and indicators to be exchanged between \emph{operational} security teams;
    \item Vocabulary for Event Recording and Incident Sharing (VERIS)\footnote{\url{http://veriscommunity.net/}} aims to be a common language for describing security incidents in a structured and repeatable manner;
    \item eXtended Abuse Reporting Format (X-ARF)\footnote{\url{http://xarf.org}}, which is used for reporting abuse to network and DNS operations;
    \item Malware Information Sharing Platform and Threat Sharing (MISP) \cite{wagner_misp_2016}, which rather focusses on operational threat intelligence such as indicators of comporomise (IoC).
\end{itemize}

Additionally, considering further literature such as the corresponding ISO and NIST documents \cite{isoiec_isoiec_2020, johnson_guide_2016} as well as especially including information about vulnerabilities in the scope of threat intelligence, the following formats and data structures are relevant:

\begin{itemize}
    \item Trusted Automated eXchange of information (TAXII) and OpenIoC \cite{maleh_big_2023};
    \item Common Security Advisory Framework (CSAF)\footnote{\url{https://www.oasis-open.org/committees/tc_home.php?wg_abbrev=csaf}} as language to exchange Security Advisories formulated in JSON;
    \item Vulnerability Exploitability eXchange (VEX), which provides the SBOM’s with transparency and an up-to-date view of the status of vulnerabilities together with the Software Bill of Materials (SBOM) as (machine-readable) inventory for software \cite{radanliev_generative_2023}; and
    \item Common Vulnerability Scoring System (CVSS)\footnote{\url{https://www.first.org/cvss/v3.1/specification-document}} for scoring vulnerabilities published as Common Vulnerabilities and Exposure (CVE)\footnote{\url{https://www.cve.org/}}, which in turn are used to assign a unique number to each vulnerability in software or hardware. The  National Vulnerability Database (NVD)\footnote{\url{tps://nvd.nist.gov/, }} receives its information from Mitre and is considered to be the primary CVE database \cite{aggarwal_study_2023}. 
\end{itemize}

\subsubsection{Integration in AI4HCTI and Data Model} 

A generic and flexible data model is being developed for AI4HCTI in order to store and process all relevant pieces of threat intelligence. Menges et al. have researched into such a Unified CTI data model, which serves as a starting point for AI4HTCI.

\subsection{Integration of Natural Language Information}
\label{label_nlp}

As shown in the introduction of this section, although health specific threat intelligence is about to be provided, this is - at least in its current stadium - only available as threat briefs and sector alerts; altogether prepared to be processed by humans. Same applies to other additional information with relevance to the sector such as news pages or sector independent threat briefs such as the ones from the European Agency for Cybersecurity, ENISA - called the ENISA Threat Landscape (ETL) \cite{european_union_agency_for_cybersecurity_enisa_2023}.

Natural language processing (NLP) as one prominent technology in Artificial Intelligence addresses exactly the challenge of retrieving and processing information provided in natural language. Basically, AI4HCTI is prepared to come up with a component, which is able to semi-automatically crawl the internet for significant sources and applies NLP to transform the relevant information into the AI4HCTI data model. This way, the information is made accessible by users through an interactive dashboard and to other components of AI4HCTI, especially the \emph{cybersecurity scoring} (\autoref{label_security_scoring}). At the core of the NLP processor a knowledge graph will be maintained, which is going to evolve over time. Knowledge graphs are relevant in the cybersecurity domain and can already assist security analysts to obtain cyber threat intelligence or discover new cyber knowledge (see \autoref{tab:knowledge_graphs} for overview) \cite{sikos_cybersecurity_2023}. 

\begin{table*} 
\caption{Comparison of prominent cybersecurity knowledge graphs (Source: \cite{sikos_cybersecurity_2023}).}
\label{tab:knowledge_graphs}
\centering
     \begin{adjustbox}{width=\textwidth,center}
        \begin{tabular}{llllll}
            \toprule
            KG & Purpose & Data model  & Implementation & Query language/software & Standard alignment\\    

            \midrule
            CSKG & General-purpose cyber-knowledge graph & RDF & OWL & SPARQL & CAPEC, CVE, CWE \\ 
            CWE-KG & Twitter data analysis & Relational data model & CSV & Log Parser or similar & CAPEC, CWE \\ 
            Live Cybersecurity KG & Security infrastructure representation & RDF & GraphDB & SPARQL & N/A \\
            Open-CyKG & Open Cyberthreat Intelligence & Uncanonicalized KG & Neo4j & Cypher & N/A \\
            MalKG & Malware threat intelligence & Tree & JSON & jsonQuery or similar & CVE \\
            SEPSES CKB & General-purpose cyber-knowledge graph & RDF & OWL & SPARQL & CAPEC, CPE, CVE, CVSS, CWE \\
            UCO & Cybersecurity standard alignment & RDF & OWL & SPARQL & STIX, CAPEC, MAEC, CWE, CVE, CVSS, Cybox, CPE, OpenIOC \\
            Vulnerability KG & Vulnerability visualization & Labeled property graph & Neo4j & Cypher &  CVE, CWE \\
            \hline
        \end{tabular}
     \end{adjustbox}
\end{table*}

\section{Outside Security Level Insights}
\label{label_deepdive2}
Besides general security laws, regulations and standards like NIST SP~800-53~\cite{NISTSO80053}, ISO~27001~\cite{ISO27001:2022}, and the General Data Protection Regulation (GDPR)~\cite{EuropeanParliament2016}, the healthcare sector is due to their critical operation and the handling of sensitive health information also subject to the Health Insurance Portability and Accountability Act (HIPAA)~\cite{hipaa}, the Directive on Security of Network and Information Systems (NIS2)~\cite{directiveeu2022/2555oftheeuropeanparliamentandofthecouncil2022} in Europe, and the National Infrastructure Protection Plan (NIPP)~\cite{u.s.departmentofhomelandsecurity2013} in the Unites States.

\begin{figure}
\centering
    \begin{tikzpicture}[text centered]
        \begin{scope}[blend group = soft light]
            \fill[red!30!white]   ( 90:1.2) circle (2);
            \fill[green!30!white] (210:1.2) circle (2);
            \fill[blue!30!white]  (330:1.2) circle (2);
        \end{scope}
        \node [font=\small,text width=2.5cm] at ( 90:2)    {Internet Scanning and Incident Databases};
        \node [font=\small,text width=1cm] at ( 210:2)   {OSINT};
        \node [font=\small,text width=1cm] at ( 330:2)   {Vulnerability Scans};
        \node [font=\small, text width=1cm] {Outside Analysis};
    \end{tikzpicture}
    \caption{Data source categories of the outside analysis.}
    \label{fig:scanning_methods}
\end{figure}
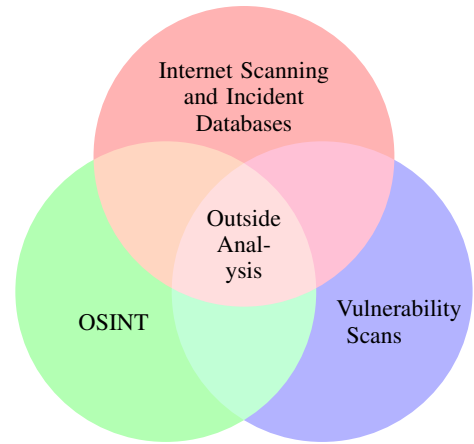

Those regulations, especially the three last mentioned, emphasize the necessity to put a strong focus on risk management, monitoring and security reporting. In order to meet the requirements, healthcare institutions need to inspect and analyze their cyber-security environment continuously. Security assessments can be performed from inside an organization, e.g. via security incident and event management systems (SIEM), intrusion detection systems (IDS), endpoint detection and response (EDR) systems, network scanners, and vulnerability scanners. These types of assessments correspond to the view of an cyber-security team within an organization. As the solution to be developed within this project does not sit within a healthcare institution, an analysis from the outside is performed. However, currently, there is no clear definition in literature, what an \textit{outside analysis} entails. Within this project an outside analysis is defined as an examination of information that can be gathered from the three different sources shown in \autoref{fig:scanning_methods} and mentioned below:
\begin{itemize}
    \item \textit{Internet Scanning and Incident Databases};
    \item \textit{Open Source Intelligence (OSINT)};
    \item \textit{Open Port and Vulnerability Scans}.\footnote{Note: Those scans will only be performed for healthcare organization, that agreed that those scans can be performed from outside their network}
\end{itemize}

This chapter's focus is on gathering security insights of specific healthcare institutions to enrich the healthcare sector specific vulnerability and threat intelligence data collected in \autoref{label_deepdive1}. In the following, \autoref{label_collectionmethods} outlines examples of information sources. In \autoref{label_limitations_benefits}, the limitations and benefits of the outside scan for this project are discussed, and finally, \autoref{label_AIforCapturing} describes how AI can improve the data capturing efficiency in this project.

\subsection{Proposed Information Sources}
\label{label_collectionmethods}
Scanning the internet for security purposes is nothing new and has been discussed in various publications in the past. For example, Durumeric et al.~\cite{182948} describe an architecture for an efficient IPv4-address space scanner considering offensive and defensive security implications. Moreover, industries have been leveraging the scanning capabilities to predict potential security incidents~\cite{8567980} in their own company or in the that of third parties they are reliant on. The following sections give brief insights on collection methods from databases and the resulting data to be further processed in \autoref{label_security_scoring}.

\subsubsection{Internet Scanning and Incident Databases}
One data source for the AI4HCTI platform is the Censys database developed by Durumeric et al.~\cite{10.1145/2810103.2813703}. The authors provide different methods to access their data, e.g. via web-based API, REST API, or downloadable raw dataset. Censys regularly scans sixteen protocols in the internet and delivers a structured and recent picture of hosts, their configurations and vulnerabilities. If combined with a reverse Whois lookup, e.g. using OWASP AMASS\footnote{\url{https://github.com/owasp-amass/amass}}, or the advanced website and domain attribution by Sebasti\'{a}n et al. ~\cite{10.1145/3627106.3627190}, the security level of health organizations worldwide or in certain geographies can be investigated. One use case Durumeric et al.~\cite{10.1145/2810103.2813703} focused on in their publication, were the scans of publicly accessible Industrial Control Systems (ICS). Analog to that, the AI4HCTI analysis can be done for OT in hospitals, e.g. medical devices like magnetic resonance imaging scanners or devices that support facility management, e.g. air conditioning or energy management. Furthermore, unintentionally externally exposed hosts, forgotten patches, host that use harmful SSLv3 protocols or sites that provide certificates signed with outdated SHA-1 modification check algorithms can be discovered. 
Another potential data source are incident reports, for example the Verizon annual Data Breach Investigations Reports (DBIR)~\cite{Verizon_Business_2024}. Incident reports over time can enrich the outside analysis to better predict trends and win insights about incident occurrence frequency.

\subsubsection{Open Source Intelligence}
\label{label_OSINT}
Further information can be collected via unstructured OSINT sources. Williams and Blum describe OSINT as a source for information that if viewed in greater context might provide significantly more insights than viewed individually~\cite{williams2018defining}. An example that highlights the importance of the context in cybersecurity is the study of Shu et al.~\cite{shu2018understanding}. They collected and analyzed sentiments of social media posts in a period of four weeks and found that a correlation between cyber incidents and public discussions on twitter exists. It would however be coincident, if a single observed post would have the same correlation.
OSINT information can be divided into five different categories and can entail multiple types of data. \textit{Geo-spatial Intelligence (GEOINT)} is one of those categories and is defined by Becastow and Bellafiore as the capability to determine, gather and analyze information to obtain better geographic understanding~\cite{bacastow2009redefining}. Based on this definition, examples of GEOINT in the context of cybersecurity are anomalies in network traffic of certain regions or identification of locations that might be more vulnerable than others. As the health sector is subject to different laws and regulations dependent on the geography in the world, GEOINT is a powerful source of information for measuring impact of security events in the industry. Another category of OSINT to be considered in this project is \textit{Social Media Intelligence (SOCMINT)}. This category entails the acquisition of information from social media (e.g. blogs, news, posts, deep web, or dark web forums) to be used as indicators for a certain research object. In the field of cybersecurity, it can for example be used for the detection, characterization, and future prediction of cyber events~\cite{10.1145/3132847.3132866}. Especially in the healthcare sector, where security awareness across employees is estimated to be rather low, characterizing security incidents and identifying indicators of compromise delivers an enormous benefit for organizations. Finally, Lowenthal considers \textit{Human Intelligence (HUMINT)}, \textit{Imagery Intelligence (IMINT)}, and \textit{Signals Intelligence (SIGINT)} as part of OSINT~\cite{lowenthal2001osint}. However, those categories are not considered further in this work.

\subsubsection{Open Port and Vulnerability Scans}
By being completely reliant on the Censys database in terms of open ports and vulnerabilities,  the actuality of the data is dependent on the latest scan in this database. Additionally, the IPv6 address range is not covered, and information about targeted companies can not be retrieved directly~\cite{tundis2018review}. Consequently, there might be hosts or parts of networks of healthcare institutions that are not covered by scans contained in this database or an automatic mapping to a specific company is not possible. Therefore, upon agreement, open ports and vulnerabilities of healthcare institutions being part of this project are examined via interaction-based tools such as Skipfish\footnote{\url{https://gitlab.com/kalilinux/packages/skipfish}}, Nessus\footnote{\url{https://www.tenable.com/products/nessus}}, Nmap\footnote{\url{https://nmap.org/}}, or Shodan\footnote{\url{https://shodan.io/}}.  

\subsection{Limitations and Benefits}
\label{label_limitations_benefits}
By making use of Censys and open port and vulnerability scans, challenges and limitations remain. One is, that it is difficult to automatically uncover and include outsourced services to the scanning results, that individual health organizations are liable for. In general, the significance of the information is only as big as the publicly facing footprint of the institution. There might also be not enough context to assess and evaluate the extracted information sufficiently. Looking at OSINT, further challenges can be encountered, such as the complex data handling~\cite{fleisher2008using}, finding relevant features in unstructured data and ignoring fake data~\cite{bello2016social}, as well as the consideration of ethical and legal aspects~\cite{bean2011open}.
Considering those limitations, the outside analysis is nevertheless important for this project. The outside perspective corresponds the first impression an attacker might have in her information gathering about a potential target. It can also give insights about the severity of potential attacks despite potential high uncertainty that comes with it. The method can furthermore unveil compliance violations (e.g. certain services might not be allowed to run on open ports). Due to the large amount of information over time, outside scans can easily be used to observe developments over days, months and years.
      
\subsection{AI for Capturing Information}
\label{label_AIforCapturing}
In order to handle large amount of data that can be collected from outside sources, machine learning models are deployed for selection, collection and quality assessment of information. According to Tundis et al. there are three main measures to be taken for identification of relevant information~\cite{TUNDIS2022102576}: 
\begin{enumerate}
    \item \textit{Comprehension of data relevance} includes the understanding of relevant threat and vulnerability information.
    \item \textit{Acquiring relevant data} from OSINT sources.
    \item \textit{Rating data source, features and collection models}.
\end{enumerate}
While the comprehension of relevant threat and vulnerabilities for the health sector is covered in \autoref{label_deepdive1}, in this section the acquisition of relevant institution specific information as well as the rating of data, features and collection models is performed. For the acquisition of data multiple use cases are discussed in literature. Examples of artificial intelligence methods for acquisition and rating of data are summarized in \autoref{tab:AI_acquisition_datarating}. The methods' practicability and further experimental setups will be examined for the practical implementation of this project.

\begin{table*}
\centering
\footnotesize
    \caption{Examples of AI methods for data acquisition and rating.}
    \label{tab:AI_acquisition_datarating}
        \begin{tabular}{p{3.5cm}p{10cm}p{2cm}}
            \ttfamily & \ttfamily \textbf{Method and Use Case} & \ttfamily \textbf{Publication} \\                \hline
            
            \multirow{4}{*}{AI for data acquisition} & Comparison of different algorithms for data acquisition, incl. Artificial Neural Network (ANN), Support Vector Machine (SVM), Random Forest, Decision Tree, Naive Bayes for acquisition of cyber threat information in the dark web. & ~\cite{osti_10420023} \\\cline{2-3}

             & Ensemble decision tree classifiers and Style Trees~\cite{10.1145/956750.956785} for predicting websites that might become compromised in the future. & ~\cite{soska2014automatically} \\\cline{2-3}

             & NLP and SVM for the extraction of indicators of compromise (IOC). & ~\cite{10.1145/2976749.2978315} \\\cline{1-3}

            \multirow{4}{*}{AI for data rating} & Comparison of SVM Regressor with Radial Basis Function (RBF) kernel, Random Forest Regression (RFR), Gradient Boosting Tree Regression (RBTR), Extra Tree Regressor (ETR), and Multilayer Perceptron Regressor (MPR) for  scoring of relevant sources. & ~\cite{TUNDIS2022102576} \\\cline{2-3}
            
             & Clustering with k-means can be used to detect duplicated values or to preprocess data for duplicate detection. & ~\cite{TUNDIS2022102576} \\\cline{1-3}
        \end{tabular}
\end{table*}
    
\section{Process Mapping}
\label{label_ProcessMapping}
\begin{figure}
  \centering
  \includegraphics[width=\linewidth]{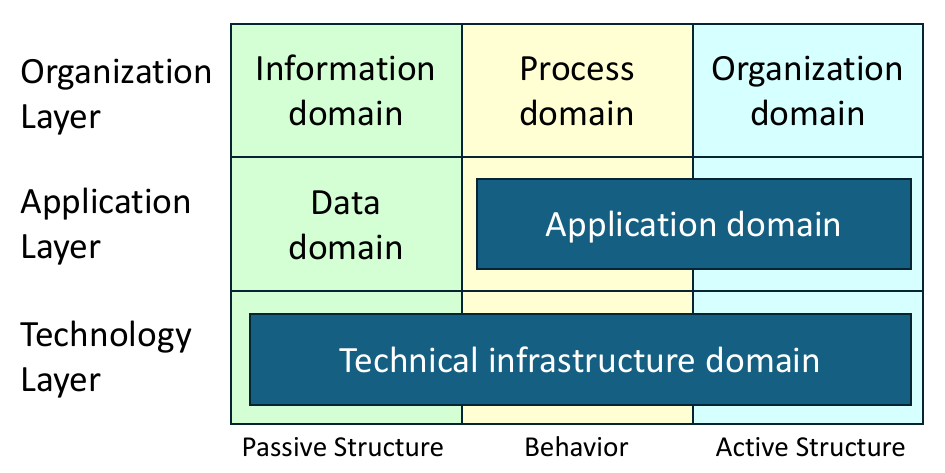}
  \caption{AI4HCTI-platform covering three layers of enterprise architecture.}
  \label{fig:AI4HCTI_EALayers}
\end{figure}
Modeling an organization is crucial to cover a three-layer view: organizational layer, application layer, and technology layer. These three layers as well as color concepts are adapted from the original ArchiMate\textregistered Meta-model (see \autoref{fig:AI4HCTI_EALayers})~\cite{Blangenois_etal_2013}. The modeling ensures a comprehensive and cohesive approach to map possible cybersecurity issues with enterprise architecture of a healthcare provider.

\textbf{Organizational layer:} This layer focuses on the operational aspects of healthcare delivery, including the treatment processes, patient flow, and administrative tasks. For the AI-enabled HCTI platform, understanding and modeling this layer helps in identifying and mapping the generic treatment process for patients. This is essential to ensure that the platform aligns with actual clinical workflows, improving both the usability and adoption of the system. By validating these processes, the platform can effectively support healthcare providers in delivering care, ensuring that the technological solutions integrate seamlessly into daily operations.

\textbf{Application layer:} This layer involves the applications and software systems that execute the business processes. In the case of the hospital information system this includes the algorithms for processing patient data, the user interfaces, and the databases that store medical information. Proper mapping and modeling of this layer are crucial to ensure that the software functionalities align with the healthcare processes they are meant to support. This layer transforms the theoretical treatment processes from the business process layer into practical, executable actions by the software systems, facilitating real-time data processing, decision support, and communication among healthcare professionals.

\textbf{Technology layer:} This layer includes the physical components—servers, computers, mobile devices, and network systems—that run the software applications. For an effective AI-driven HCTI platform, it’s critical that the hardware is included in the analysis as it supports the software layer and the overall business processes of a healthcare organization. It needs to handle large volumes of data securely and reliably, maintain uptime, and ensure fast processing speeds to support real-time medical decision-making.

Focusing on the identification, mapping, and validation of the generic treatment process in these three layers ensures that the HCTI platform is not only covering technological aspects of cybersecurity but also practically useful and directly relevant linkage to patient care process. By addressing each layer systematically, organizations can mitigate risks, reduce costs, and enable a smoother transition when implementing new technologies and reduce functional redundancy \cite{Winter_etal_2009}. 
  
\section{Cybersecurity Risk Scoring}
\label{label_security_scoring}
As cyber events in the healthcare sector cannot only have a high financial impact on institutions but can also lead to life threatening situations, it is necessary to quantify the risk in a way that decision makers can easily assess it and prioritize countermeasures continuously. Therefore, \autoref{label_DefinitionOfRisk} describes the different components of risks considered in this project, and finally, in \autoref{label_AImethodsForRiskScore} the AI methods for computing the risk score are elaborated. In both sections, the focus lies on healthcare specific aspects.

\subsection{Definition of Quantifiable Risk and Its Components}
\label{label_DefinitionOfRisk}
In order to quantify cyber risk for healthcare institutions a clear understanding of quantifiable risk in general is necessary. In this project we understand risk based on the definition of Kaplan and Garick~\cite{kaplan1981quantitative} as a set of the four following elements:
\begin{enumerate}[wide, labelwidth=!, labelindent=0pt]
    \item A \textit{scenario} describes a contingency that might materialize with a certain probability and impact.
    \item The \textit{probability} describes the chance that a specific scenario occurs and brings some uncertainty to the table.
    \item The \textit{impact} of a certain scenario is the measurable consequence of a scenario that comes true.
    \item The \textit{number of relevant scenarios} is the quantity of scenarios that have been identified to be relevant for a certain organization.
\end{enumerate}
In addition to the definition of Kaplan and Garick, in the present project one other element of risk is introduced. The probability of occurrence and the impact are dependent on the implemented measures that reduce the risk which should be considered when predicting the risk score. Therefore, the computation of the net risk takes a score for countermeasures into account. If the in \autoref{label_deepdive2} described methods discover gaps of countermeasures based on the outside analysis the countermeasure score is decreasing.
Consequently, the net risk for this project can be annotated as in \eqref{eq:risk}.
\[
R={\{s_i,p_i,m_i,c_i}\}, \;\;\; \mbox{for} \;\;\; i=1,...,n \tag{1} \label{eq:risk} 
\]
where \newline
\hspace*{0.8cm} \(R\) = risk; \newline
\hspace*{0.8cm} \(s\) = scenario; \newline
\hspace*{0.8cm} \(p\) = probability of the scenario to occur; \newline
\hspace*{0.8cm} \(m\) = impact of a certain scenario; \newline
\hspace*{0.8cm} \(c\) = countermeasure to decrease the impact / probability; \newline
\hspace*{0.8cm} \(n\) = number of relevant scenarios.

\begin{figure}
    \centering
    \includegraphics[width=1\linewidth]{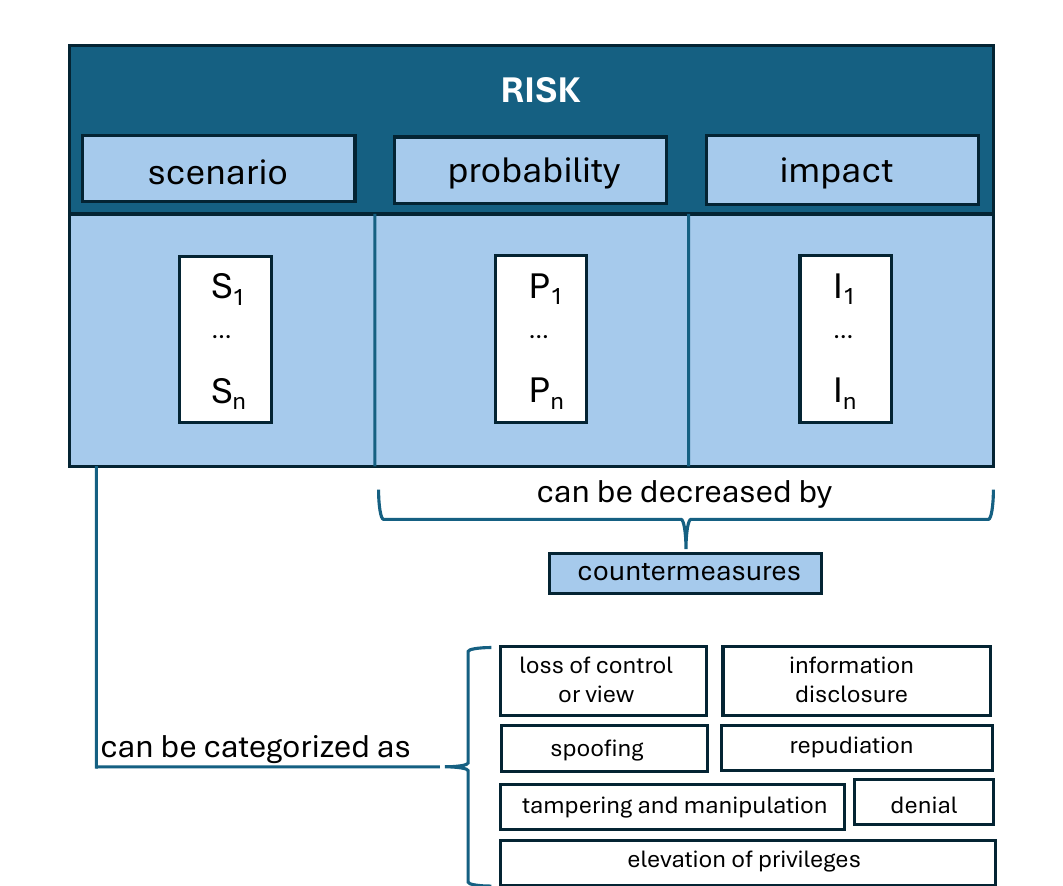}
    \caption{Elements of risk in AI4HCT.}
    \label{fig:risk}
\end{figure}

In the context of cybersecurity scoring for the healthcare sector, a starting point for defining scenarios builds the three-layer view of an organization described in \autoref{label_ProcessMapping}. For the service and application layer and the infrastructure layer, the attacker goals defined in the SANS publication of Assante and Lee ~\cite{assante2015industrial} focusing on industrial control systems (ICS) are considered and expanded to include attacker goals for medical devices and information technology (IT) as defined by the Microsoft STRIDE Threat Model\footnote{\url{https://learn.microsoft.com/en-us/azure/security/develop/threat-modeling-tool-threats\#stride-model}}. Scenarios for this project can therefore be classified as follows:
\begin{enumerate}[wide, labelwidth=!, labelindent=0pt]
    \item \textit{Loss of control or view} describes the situation, in which data in health systems is still successfully processed but devices do not behave as intended.
    \item \textit{Denial} is the category for all scenarios where control, view or safety functions of medical devices are denied. In this case it is not possible to reach the devices through the regular communication channel. Denial also includes those scenarios in which resources to communicate with and through the devices of administrative IT are exhausted. 
    \item \textit{Tampering and manipulation} entails circumstances where the control, view, sensors, or safety functions are manipulated so that false information is reported to device operators or the physical movement of devices is altered. Manipulation can also entail malicious alterations in configuration or data of administrative IT.
    \item \textit{Spoofing} describes those scenarios where the adversary takes over the identity of another user (e.g. the administrator) to increase her possibilities to access a wide range of systems, for example in order to explore where valuable data is stored and processed or where to maximize damage.
    \item \textit{Repudiation} includes all those events in which the attacker performs actions to deny her involvement in a security event (e.g. deletion of log files).
    \item \textit{Information disclosure} describes events in which an adversary accesses, steals and publishes sensitive information.
    \item \textit{Elevation of privileges} entails scenarios in which an attacker increases her access rights.
\end{enumerate}
The probability of the scenario to occur is based on two sources: OSINT and historical incident information that is collected in \autoref{label_OSINT}. The impact estimation for each scenario is based on expert interviews, OSINT for continuous monitoring, and whether an specific scenario triggers any compliance violations which would increase the impact from a financial and reputational perspective further. The platform has only limited view on whether countermeasures are implemented or not. However, the vulnerability scan in \autoref{label_collectionmethods} gives insights on a few aspects regarding the protection level of institutions. The feasibility of a scenario is given by the concrete business processes and systems of the healthcare institution. 
The elements of risk are summarized in \autoref{fig:risk}.

\subsection{AI Methods for Calculating the Risk Score}
\label{label_AImethodsForRiskScore}
In order to calculate a precise risk score, AI4HCTI makes use of AI implementations to dynamically ingest and learn from information about novel threats, vulnerabilities and outside assessments. Therefore, a feature vector is generated and the risk score for the probability is computed. For that, the approach of Bilge et al. is applied~\cite{10.1145/3133956.3134022}. They use semi-supervised learning on a random forest classifier to predict the probability of a device being infected in the future. In this project the aim is to predict the probability of occurrence as well as the impact of relevant scenarios related to healthcare institutions. For that, it has to be taken into account, that in many healthcare organizations, and in hospitals specifically, cybersecurity duties are split between different departments. One example is that the IT department typically deals with any issue that arises out of the network and administrative IT, while the medical device department is responsible for technical challenges of medical devices. When calculating the risk score, both aspects need to be considered. The probability score for a certain scenario to come true for a feature vector $X_i$ can be notated as in \autoref{eq:probability}.

\[
P_i=P(s_i,X_i) \tag{2} \label{eq:probability} 
\]
The impact score of the platform will consist of three classes, which are \textit{low impact}, \textit{medium impact} and \textit{high impact}. While the meaning of those categories is defined together with industry experts, the calculation of the impact is based on the approach of Palsson et al.~\cite{palsson2020}, who have used a random forest classifier to predict the impact class for certain scenarios.

\begin{figure}
    \centering
    \includegraphics[width=0.9\linewidth]{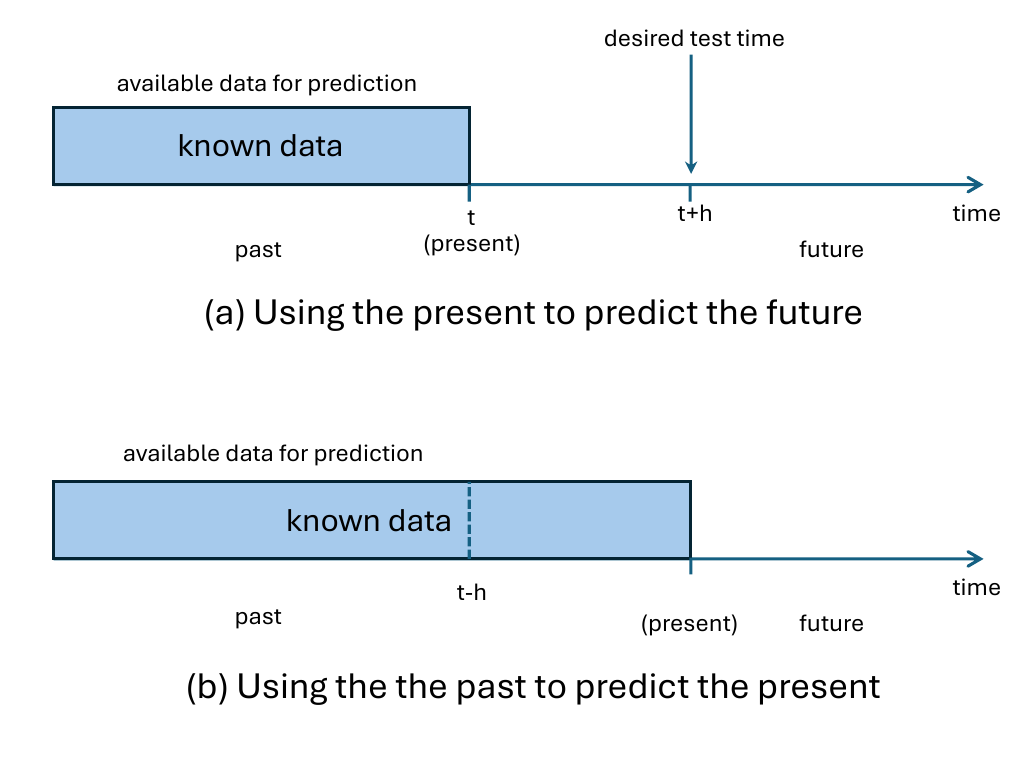}
    \caption{Using the factor time for predicting the present in training and the future in production as proposed in ~\cite{soska2014automatically}.}
    \label{fig:AI_forPrediction}
\end{figure}

As the future is unknown, it is difficult to evaluate a model that predicts the future probability and impact. Therefore, the approach of Soska and Christin~\cite{soska2014automatically}, as shown in \autoref{fig:AI_forPrediction} is considered for training the risk scoring model. The method uses data from the past to predict the present in the training and test phase and does then use further data in time to predict the future. 

\subsection{Evaluation Method}
\label{label_evaluation}
In order to evaluate the risk scoring and the result of the AI4HCTI platform respectively, different metrics are applied in order to calculate the mean calibration error between the forecasted risk and the actual observed
risk. These metrics include the mean absolute value (MAE), mean squared error (MSE), root mean squared error (RMSE), and the mean absolute percentage error (MAPE). Additionally, once a cyber incident that was previously predicted hits a certain technical healthcare setting, the previously computed score is evaluated through expert judgement again.

\section{Discussion and future work}
AI4HCTI presents the first platform of its kind that uses an assemble of AI methods for collection, assessment, extraction, and analysis of relevant cybersecurity information in the healthcare sector. While there is a placeholder for AI-enabled recommendations in our platform architecture, further work has to focus on the specific implementation of those recommendations. Furthermore, the second generation of the platform is meant to consider the analysis of platform user interaction to automatically capture hot topics that are of interest to the healthcare sector. Those focus topics can be fed back into the learning algorithms to align the focus of scraping OSINT information. 
In addition, the initial data model described in \autoref{label_prestructured}, has to be further developed in order to integrate additional formats. On top, it will need to be discussed, whether additional adaptions are necessary for integrating the threat intelligence, which shall be extracted through processing natural language documents (\autoref{label_nlp}). Indicators and threat intelligence events were explicitly dropped for the unified CTI data model of Menges; this hypothesis has to be revisited in the context of AI4HCTI; if relevant, the data model has to be adopted in this regard. Future research could also use internal, live security monitoring data from intrusion detection and response systems to enrich CTI within the AIHCTI. This would lead to a holistic view on how to search for emerging threats within the networks of healthcare institutions.

\bibliographystyle{IEEEtran}
\bibliography{arxiv}

\end{document}